\documentclass[manuscript,prl,showpacs,preprintnumbers,amsmath,amssymb]{revtex4}
\usepackage[english]{babel}
\usepackage{ucs}
\usepackage[utf8x]{inputenc}
\usepackage[dvipdf]{graphicx}
\usepackage{amssymb}

\newcommand{\sw}[0]{ {$SW$(55-77)} }

\begin{document}

\title{Stone-Wales--type transformations in carbon nanostructures
  driven by electron irradiation}

\author{J. Kotakoski$^{1,}$\footnote{Corresponding author, email:
    {\texttt jani.kotakoski@iki.fi}},
  J.~C. Meyer$^{2,}$\footnote{Present address: University of Vienna,
    Department of Physics, 1090 Wien, Austria}, S. Kurasch$^{2}$,
  D. Santos-Cottin$^{1}$, U. Kaiser$^2$ and
  A.~V. Krasheninnikov$^{1,3}$}

\affiliation{$^1$ Department of Physics, University of Helsinki,
  P.O. Box 43, 00014 Helsinki, Finland\\
  $^2$
  Electron microscopy of materials science, University of Ulm,
  Germany\\
  $^3$ Department of Applied
  Physics, Aalto University, P.O. Box 1100, 00076 Aalto, Finland\\
}

 \date{\today}


% 68.37.Og High-resolution transmission electron microscopy (HRTEM)
% 61.48.Gh Structure of graphene
% 81.05.ue Graphene (81. Materials science)
% 64.70.Nd Structural transitions in nanoscale materials
% 61.80.Fe Electron and positron radiation effects
% 31.15.es Applications of density-functional theory (e.g.,
%          defect formation)

\begin{abstract}

Observations of topological defects associated with Stone-Wales--type
transformations ({\it i.e.}, bond rotations) in high resolution
transmission electron microscopy (HRTEM) images of carbon
nanostructures are at odds with the equilibrium thermodynamics of
these systems. Here, by combining aberration-corrected HRTEM
experiments and atomistic simulations, we show that such defects can
be formed by single electron impacts, and remarkably, at electron
energies below the threshold for atomic displacements. We further
study the mechanisms of irradiation-driven bond rotations, and explain
why electron irradiation at moderate electron energies ($\sim$100~keV)
tends to amorphize rather than perforate graphene. We also show via
simulations that Stone-Wales defects can appear in curved graphitic
structures due to incomplete recombination of irradiation-induced
Frenkel defects, similar to formation of Wigner-type defects in
silicon.

\end{abstract}

\maketitle

\section{Introduction}

Stone-Wales defect~\cite{Thrower69,Stone1986} -- {\sw} -- is the
simplest example of topological disorder in graphene and other
$sp^2$-hybridized carbon systems. It can be formed by rotating a C-C
bond by $90^\circ$ with regard to the midpoint of the bond -- referred
to as the SW transformation -- so that four hexagons are turned into
two pentagons and two heptagons. This defect has received considerable
amount of
attention~\cite{Ma2009a,Ertekin2009,Zhou2003,Li2005,Miyamoto2004,Suenaga2007},
because it has the lowest formation energy among all intrinsic defects
in graphenic systems, and because it presumably plays an important
role in plastic deformation of carbon nanotubes (CNT) under
tension~\cite{Huang2006} by relieving strain~\cite{Nardelli1998}. It
can also act as a source for dislocation
dipoles~\cite{Ertekin2009,Ding2007}.

Regardless of being the lowest energy defect in
graphene~\cite{Banhart2011} and other $sp^2$-hybridized carbon
nanostructures, the {\sw} needs about 5~eV to appear in
graphene~\cite{Ma2009a,Li2005}, and 3--5~eV in CNTs with a diameter
above 1~nm~\cite{Ertekin2009,Zhou2003}, which should lead to a
negligible equilibrium concentration of such defects at room
temperature. However, recent advances in HRTEM have allowed the
identification of such defects in
graphene~\cite{Meyer2008,Kotakoski2011} and
CNTs~\cite{Suenaga2007}. Moreover, SW transformations play an
important role in the response of graphene to electron
irradiation~\cite{Kotakoski2011,Song2011}, leading to changes in the
morphology of vacancy-type defects~\cite{Lee2005b} and to their
migration. Such changes are equally surprising, because the barrier
for bond rotation is about 5 eV~\cite{Li2005,Cretu2010}, which should
exclude thermal activation as a cause for SW transformation at room
temperature during experimentally relevant time scales. Regarding
irradiation effects, previous simulations~\cite{Yazyev2007} showed
that an energy of $\sim 30$~eV must be transferred to a C atom in
graphene in the in-plane direction for a bond rotation to occur. Also
this cannot explain the frequently observed SW transformations under
the usual TEM imaging conditions, since with typical acceleration
voltages ($\lesssim 300$~kV) the transferred kinetic energy in the
direction almost perpendicular to the electron beam will remain
significantly below 10~eV.

Here, by combining aberration-corrected (AC-) HRTEM with atomistic
computer simulations, we show that topological defects associated with
the SW transformation can be formed in $sp^2$-hybridized carbon
nanostructures by impacts of individual electrons at energies even
{\it below} the threshold for a carbon atom displacement.  We further
study in detail the mechanisms of irradiation-driven bond rotations
for pre-existing vacancy-type defect structures and how they transform
and migrate due to electron impacts.  At the same time we explain why
electron irradiation at moderate energies ($\sim 100$~keV) tends to
rather amorphize~\cite{Kotakoski2011} than perforate graphene. We also
show via simulations that the {\sw} can appear in curved graphitic
structures due to ``incomplete'' recombination of irradiation-induced
Frenkel defects, reminiscent of the formation of Wigner-type defects
in silicon~\cite{Tang1997}.

\section{Methods}

\subsection{Experimental Methods}

Graphene membranes used in our experiments were prepared by mechanical
exfoliation of graphite on Si/SiO$_2$ substrates and transfer of the
resulting graphene sheets onto TEM grids as described
previously~\cite{Meyer2008a}.  For TEM experiments we used an FEI
{\scshape Titan} $80-300$ equipped with an image-side aberration
corrector, operated at 80~kV. The spherical aberration was set to
15~$\mu$m and images were recorded at Scherzer defocus. The extraction
voltage of the source was reduced to 2~kV and the condensor lens C2
was switched off in order to minimize the energy spread. Under these
conditions, dark contrast in the images can be directly interpreted in
terms of the atomic structure. Image sequences were recorded on a CCD
camera with exposure times of 1~s and intervals of approximately 2~s.

\subsection{Computational Methods}

We carried out atomistic computer simulations based on the
spin-polarized density functional theory (DFT) implemented in the
plane wave basis set {\scshape vasp} code~\cite{Kresse1996}. The
projector augmented wave potentials~\cite{Blochl1994} were used to
describe the core electrons, and the generalized gradient
approximation of Perdew, Burke and Ernzernhof~\cite{Perdew1996} for
exchange and correlation. We included plane waves up to a kinetic
energy of 300~eV. The $\mathbf{k}$-point sampling of the Brillouin
zone was performed using the scheme of
Monkhorst-Pack~\cite{Monkhorst1976} for the periodic dimensions.
Structure relaxation calculations were combined with molecular
dynamics (DFT-MD) simulations with a lower kinetic energy threshold
and fewer $\mathbf{k}$-points.

Due to the high computational cost of the DFT-MD method, only a few
simulations were carried out at this level. Whenever statistics needed
to be gathered, we calculated the forces using the non-orthogonal
DFT-based tight binding (DFTB) method~\cite{Frauenheim2002}. The main
results were checked against DFT-MD. In total, we carried out $\sim
27.000$ dynamical DFTB-MD simulations. The simulated structures
consisted of 120--200 atoms and were fully optimized. For the
displacement threshold simulations, one of the atoms was assigned a
kinetic energy $T$ with the initial velocity vector pointing to a
preselected direction. The initial temperature of the system was set
to 5~K, although we observed no differences when carrying out the
simulations for initially stationary atoms. Displacement threshold
$T_d$ (minimum kinetic energy required to eject the atom) was found to
be 22.50~eV, in a good agreement with earlier DFTB
results~\cite{Krasheninnikov2005b,Zobelli2007a}. It is also close to
the DFT value (22.03~eV)~\cite{Kotakoski2010}. For the annihilation
simulations, various system temperatures were studied (500--1200~K)
both to fasten the migration of the adatoms and to understand the
effect of an elevated temperature (as will be mentioned below).

\section{Results and Discussion}

\subsection{Stone-Wales Defects due to Single Electron Impacts}

\begin{figure}
\includegraphics[width=\textwidth]{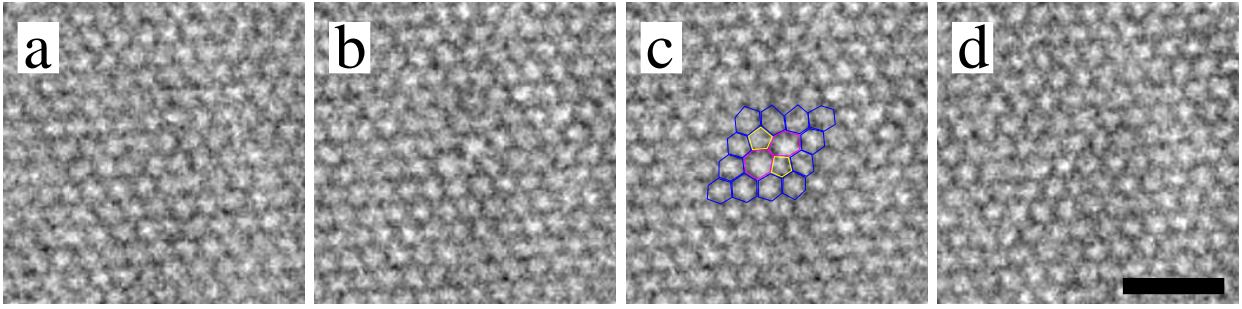}
\caption{HRTEM images of an initially pristine graphene sample (a),
  followed by a {\sw} (b). Frame (c) shows the same image with a
  structure overlay. {\sw} disappears in the following frame
  (d). Scale bar is 1~nm. (See also video S1 in Ref.~\cite{Suppl}.)}
\label{px:SWExp}
\end{figure}

We begin the presentation of our results with the description of
experimental observations of {\sw} in HRTEM images. Several long image
sequences, typically containing hundreds of images from clean and
initially defect-free graphene membranes, were recorded at
80~kV. Occasionally, {\sw} defects appear in individual exposures, as
in the example shown in Figure~\ref{px:SWExp}b (Figure~\ref{px:SWExp}c with
structure overlay). Remarkably, in most of the observed cases,
isolated {\sw} appeared in pristine graphene for one 1~s exposure,
only to disappear in the following frame. Hence, the lifetime of this
defect under the 80~kV electron beam in terms of irradiation dose is
of the order of $10^{7}~e^{-}/\textrm{nm}^2$, the dose used for a
single exposure.

\begin{figure}
\includegraphics[width=.8\textwidth]{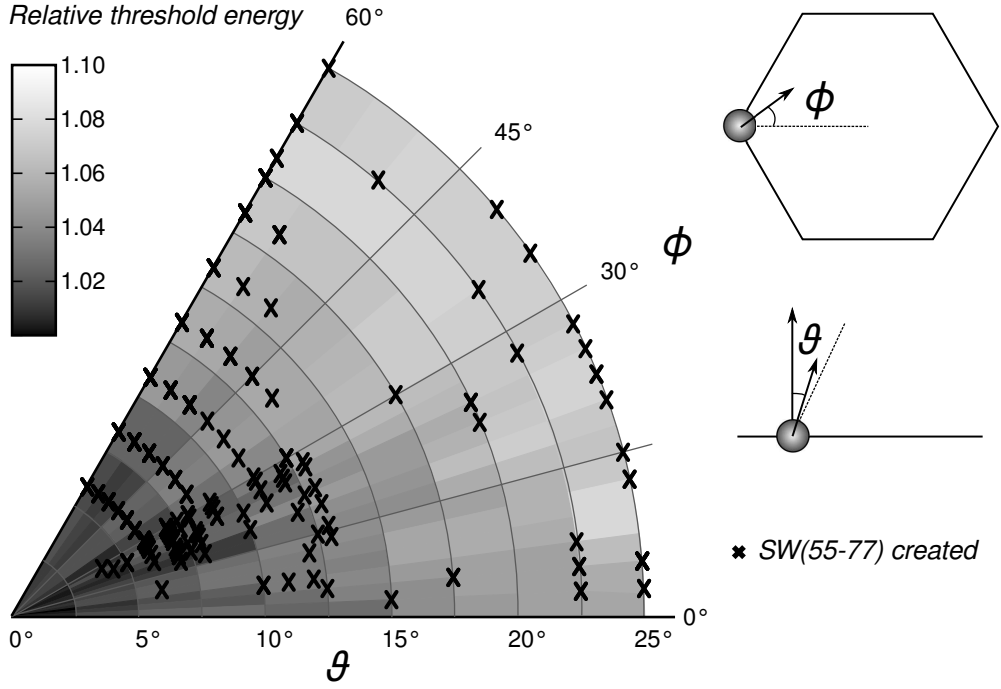}
\caption{Relative displacement threshold
  $T_d^{\phi,\theta}/T_d^{\theta=0}$ as a function of the displacement
  space angle ($\theta$, $\phi$). Crosses mark the angles for which we
  observed impact-induced SW transformations.}
\label{px::sect}
\end{figure}

To understand the appearance and disappearance of {\sw}, we carried
out atomistic simulations for individual displacement events under the
electron beam. After calculating $T_d$, we extended this calculation
to all in-plane ($\phi$) and out-of-plane angles in the range $\theta
\in [0^\circ, 25^{\circ}]$, Figure~\ref{px::sect}. Displacements with
$\theta>25^\circ$ would result in transferred kinetic energies of more
than 2~eV below $T_d$ for an electron beam even at 120~keV. Since the
displacement threshold increases for increasing $\theta$, it is
unlikely that this restriction would lead us to miss any significant
electron beam-induced structural changes, especially for electron
energies similar to those used in this study (80~keV). The calculated
displacement thresholds are shown in Figure~\ref{px::sect} in a
relative scale along with the space angles for which we observed the
formation of {\sw}.

It is evident from Figure~\ref{px::sect} that the SW transformation is a very
likely event at displacement angles slightly away from the graphene
plane normal ($\theta\geq 2.5^\circ$). The transferred kinetic
energies ($T$) required for this process are below the displacement
threshold for the corresponding $\phi$ and $\theta$
($T_d^{\phi,\theta}$) since no actual removal of the recoil atom is
required for the bond rotation to occur. Typically, $T\approx
T_d^{\phi,\theta} - 1$~eV resulted in the \sw formation, although for
some space angles even $T \approx T_d^{\phi,\theta} - 2$~eV was
enough. The probability for {\sw} formation is particularly high for
certain space angles, which is related to different mechanisms of SW
transformation, as described below.

\begin{figure}
\includegraphics[width=.8\textwidth]{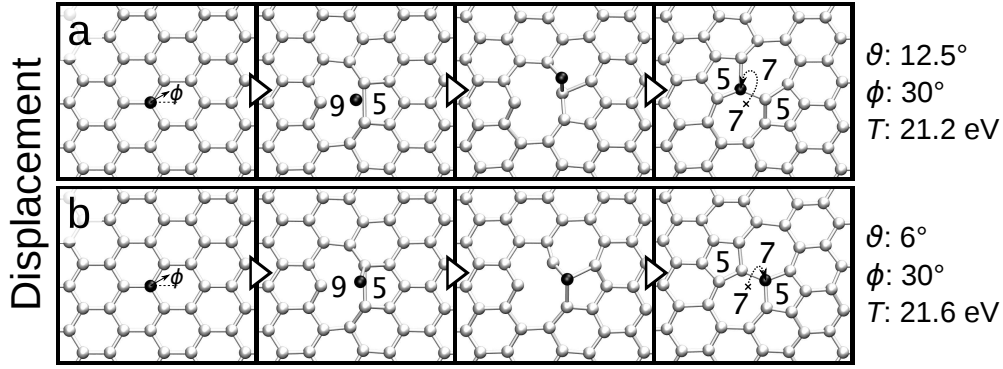}
\caption{Typical SW transformation processes for low out-of-plane
  displacement angles ($\theta \leq 25^\circ$): (a) The ``circle''
  process, and (b) the ``nudge'' process. The black spheres correspond
  to the recoil (displaced) atoms, and the mono-vacancy structure is
  highlighted with numbered carbon rings in the second panels. The
  last panel shows schematically the route of the displaced atom. (See
  also videos S8 and S9 in Ref.~\cite{Suppl}.)}
\label{px::swprocs}
\end{figure}

The above-presented result is in clear contrast with the earlier
simulation results for graphite~\cite{Yazyev2007} where no {\sw}
formation was observed for low $\theta$. This discrepancy is caused by
the neighboring graphene planes in the case of graphite: The displaced
atom gets attached to the adjacent layer and does not therefore
initiate a bond rotation. In Figure~\ref{px::swprocs} we show the two
processes which account for the majority of the SW transformations
observed during our simulations. In the ``circle'' process
(Figure~\ref{px::swprocs}a), the displaced atom circles around its neighbor,
whereas in the ``nudge'' process (Figure~\ref{px::swprocs}b) it nudges the
neighbor to cause the bond rotation. Note that the example cases are
for the same $\phi$ and almost same $T$, but for different
$\theta$. The resulting process for each displacement is an interplay
of all three variables ($T$, $\theta$, $\phi$).

Similar mechanisms also exist for CNTs. However, as two new parameters
(tube diameter and chirality) should be introduced for quantitative
analysis of SW transformation, we did not study this process in
nanotubes at length due to unreasonably high computational cost.

\subsection{Annihilation of Vacancies and Adatoms with Possible Formation of {\sw}}

Since we frequently observed formation of vacancy--adatom pairs
(adatoms play the role of interstitials in graphene and CNTs) in our
simulations of electron impacts onto graphene and earlier in
CNTs~\cite{Krasheninnikov2005b}, we also explored another possible
mechanism of {\sw} formation, which is based on ``incomplete''
annihilation of a Frenkel defect. This study was motivated by the
peculiarities of the recombination of such a defect in bulk
silicon. In that covalently-bonded material the recombination can give
rise to either annihilation of the defect and restoration of the
perfect crystal lattice, or to a Wigner-type
defect~\cite{Tang1997}. Such topological defects are imperfections in
the crystal lattice with the locally ``correct'' number of atoms (as
opposed to vacancies and interstitials), with the atomic configuration
separated from the perfect structure by a finite potential
barrier. Such defects are deemed to also exist in
graphite~\cite{Ewels2003,Telling2003}, and {\sw} can clearly be
classified into this group.

\begin{figure}
\includegraphics[width=.8\textwidth]{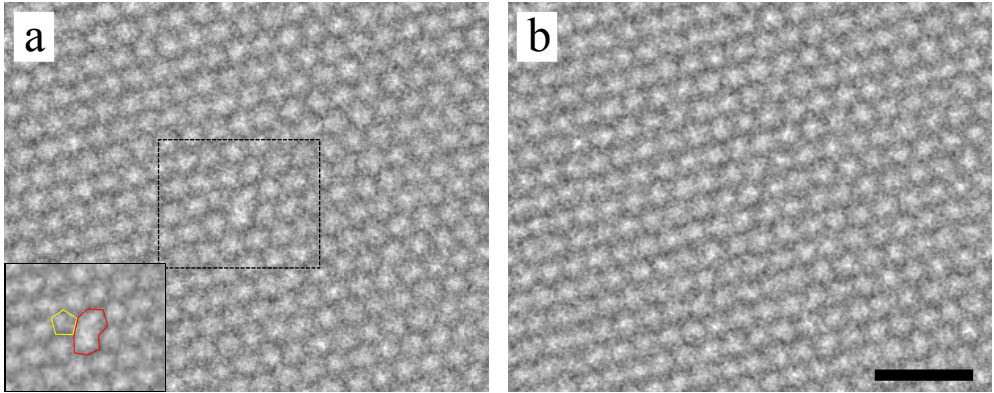}
\caption{Experimental image sequence of a vacancy annihilation. Panel
  (a) shows a $V_1$(5-9) mono-vacancy, with an overlay in the inset,
  and (b) shows the same region in a later exposure, with no defect
  visible. Scale bar is 1~nm. (See also video S7 in Ref.~\cite{Suppl}.)}
\label{px::expannih}
\end{figure}

While carbon adatoms on graphene~\cite{Lehtinen2003,Erni2010} and
CNTs~\cite{Gan2008,Krasheninnikov2006} (especially those inside
nanotubes) are mobile at room temperature, they can easily find
vacancies in the system and annihilate. Indeed, we occasionally
observed disappearance of vacancies in HRTEM image
sequences. Figure~\ref{px::expannih} shows an example of a mono-vacancy that
disappears during observation. This proves that mobile carbon atoms
are present under our experimental conditions, and may recombine with
vacancy-type defects. However, we never noticed the creation of a
{\sw} after an observed mono-vacancy.

\begin{figure}
\includegraphics[width=.8\textwidth]{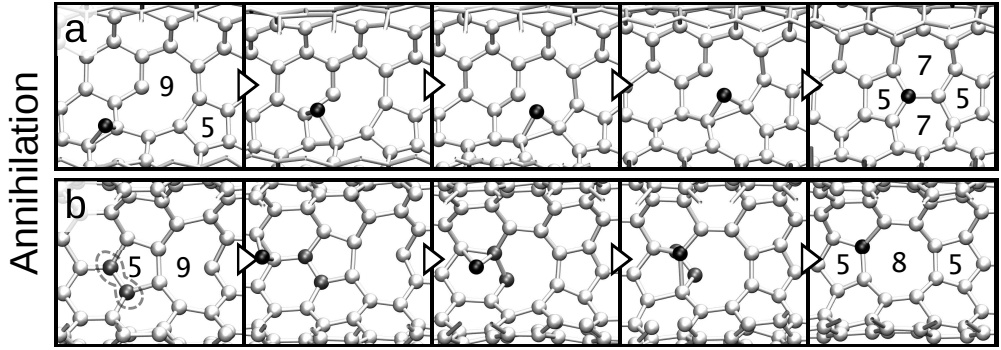}
\caption{(a) {\sw} creation process as an outcome of a vacancy--adatom
  annihilation in a (6,6) armchair nanotube. (b) Carbon dimer
  sputtering process in a (10,0) zigzag nanotube. The black spheres
  stand for adatoms and the gray ones denote the dimer to be sputtered
  [marked in the first panel of (b)]. The nanotube structures are
  shown from inside the tube. The tube axis is in the horizontal
  direction. (See also videos S10 and S11 in Ref.~\cite{Suppl}.)}
\label{px::swannih}
\end{figure}

To simulate the annihilation process, we created nearby Frenkel
defects (separated by a few {\AA}ngstr\"oms) in a graphene layer and
small (6,6) zigzag and (10,0) armchair CNTs (radii $r\approx
4.1$~{\AA} and $\sim 3.9$~{\AA}, respectively). We then heated the
structures and collected statistics on the evolution of each system by
running dynamical atomistic simulations at various temperatures
(500--1200~K). Three possible outcomes emerged from the simulations:
(1) perfect annihilation to the pristine structure (similar to the
experimental images in Figure~\ref{px::expannih}), (2) formation of a {\sw}
(Figure~\ref{px::swannih}a) and, surprisingly, (3) sputtering of a C$_2$
dimer with a remaining reconstructed di-vacancy $V_2$(5-8-5)
(Figure~\ref{px::swannih}b).

For graphene, we always observed perfect annihilation in accordance
with the experiments. However, if the adatom in graphene was placed on
top of one of those two bonds in the nine-membered carbon ring which
are right next to the pentagon, a {\sw} was spontaneously formed
without an energy barrier. Thus {\sw} may also form in graphene due to
recombination of Frenkel defects, but the probability for this process
must be much lower than in CNTs.

For (10,0) CNT, we obtained perfect annihilation in approximately 54\%
of cases, {\sw} was formed in approximately 34\% of cases and dimer
sputtering occurred in approximately 12\% of cases. For (6,6) CNT, the
values were 53\%, 42\% and 4\%, respectively. With increasing
temperature the probability to sputter a dimer showed a slight
tendency to increase. We also ran the calculations for a (8,8) CNT
($r\approx 5.4$~{\AA}) at 800~K in order to estimate the curvature
dependency of the results. The values did not significantly differ
from those for the (6,6) CNT, except for a somewhat increased tendency
to perfect annihilation and decreased sputtering (with probabilities
of 56\% and 2\%, respectively).

The formation energy of a spatially separated Frenkel defect in
graphene within the DFTB model is approximately
11.1~eV. Transformation to form SW defect from this initial setup
leads to an energy gain of 5.4~eV. By sputtering a dimer, graphene
would instead gain 0.4~eV. Hence, all three observed outcomes are
energetically reasonable also for graphene. For nanotubes, the
formation energies of both {\sw} and a di-vacancy are lowered due to
the curvature and stronger C-C bonds at
pentagons~\cite{Krasheninnikov2006}. The corresponding energy gains
are also higher, which can explain why the probability for {\sw}
defect formation and dimer evolution is higher in curved carbon
nanostructures. The actual energies depend on the local curvature. It
is also plausible that the dimer evaporation process plays a role in
shrinking fullerenes under electron irradiation~\cite{Huang2007}.

\subsection{Stone-Wales Transformations in Vacancy-Type Defects}

{\sw} defect represents the elementary case of a topological change in
the graphene structure, {\it i.e.}, a single bond rotation in the
otherwise perfect structure.  More abundant, however, are changes in
the atomic configuration through bond rotations in the reconstructed
vacancy defects, as recent experiments
indicate~\cite{Kotakoski2011,Song2011}. In presence of a
(multi-)vacancy, the atomic configuration of a defect can be
transformed between different metastable structures via bond
rotations. It was presumed~\cite{Kotakoski2011} that such SW
transformations of vacancy-type defects would be stimulated by
electron impacts, but the actual atomistic mechanism has not been
hitherto unraveled.

In order to get microscopic insight into irradiation-stimulated bond
rotations near vacancy-type defects in graphene, we carried out a set
of experiments and dedicated simulations aimed at assessing the
probability of SW transformations in the defect structures. In our
experiments, we initially generated ``defective'' graphene by brief
150~kV electron irradiation, and then recorded image sequences of
di-vacancies using 80~kV AC-HRTEM.  As observed
previously~\cite{Kotakoski2011}, the vacancies can transform between
different configurations under the influence of the 80~kV electron
beam. Moreover, the di-vacancies migrate and transition between the
different reconstructed configurations via SW transformations,
typically until they cluster into larger defects (see videos S5 and S6
in Ref.~\cite{Suppl}).

\begin{figure}
\includegraphics[width=.8\textwidth]{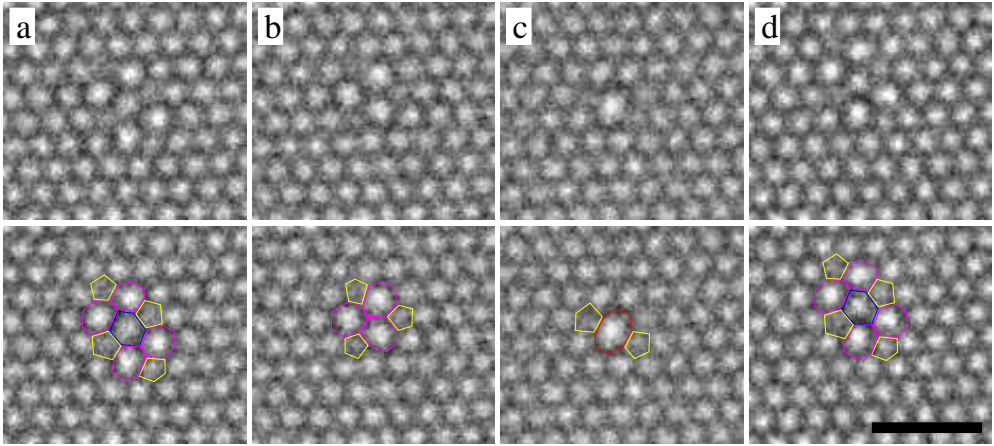}
\caption{Experimental images illustrating SW transformations in the
  atomic structure of di-vacancies and migration of these defects.
  Panels (a-d) show sequential HRTEM images of the same defect, in the
  lower row with structure overlay. (a) $V_2$(5555-6-7777)
  transforming into $V_2$(555-777) (b), and $V_2$(5-8-5) (c). Each of
  these transitions can be explained by a single bond rotation. In a
  later frame (d), the defect is again a $V_2$(5555-6-7777), but
  shifted by one lattice parameter. Scale bar is 1~nm. (See also
  videos S2--4 in Ref.~\cite{Suppl}.)}
\label{px::mvacexp}
\end{figure}

In Figure~\ref{px::mvacexp}, we show an example of a di-vacancy defect that
transforms between different reconstructed shapes [$V_2$(5-8-5),
  $V_2$(555-777), $V_2$(5555-6-7777)] under the electron beam. The
changes in the atomic structure of these di-vacancy configurations can
be described by SW transformations at the defect. Moreover, multiple
transformations allow migration of the di-vacancy. Similar to {\sw}
defect formation discussed above, the activation energy for these
transitions is far too high to allow a thermally activated process
with an observable rate at room temperature. Hence, on the basis of
the observations, the activation energy for the transition must be
provided by the electron beam.

\begin{figure}
\includegraphics[width=.8\textwidth]{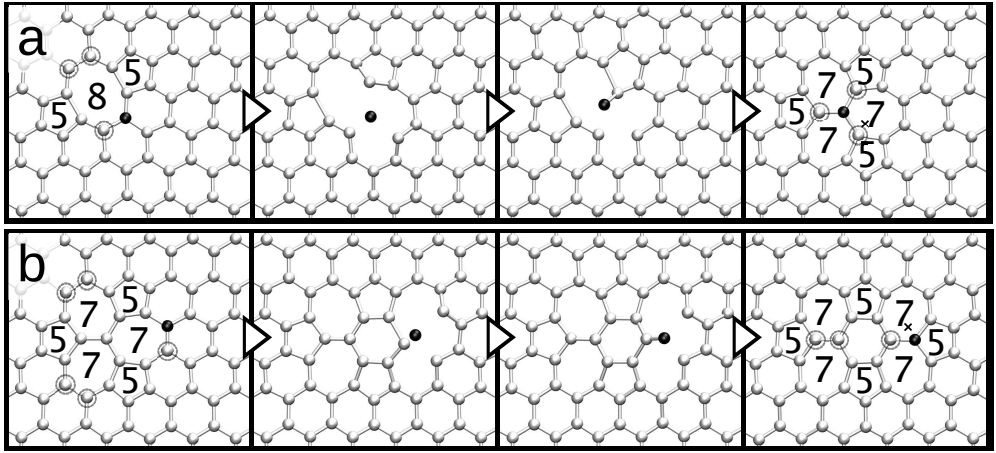}
\caption{Changes in the atomic structures of reconstructed di-vacancy
  defects through SW transformations upon atom displacements
  perpendicular to the graphene sheet ($\theta=0^\circ$); (a)
  $V_2$(5-8-5)$\rightarrow V_2$(555-777) and (b)
  $V_2$(555-777)$\rightarrow V_2$(5555-6-7777). The black spheres
  indicate the recoil (displaced) atoms. Structurally equivalent atoms
  to the displaced ones are marked with circles in first panels. In
  the last panels, the circled atoms are those which could cause a
  backward transformation upon displacement in addition to the recoil
  atom. (See also videos S12--14 in Ref.~\cite{Suppl}.)}
\label{px::mvacsim}
\end{figure}

In order to confirm that the SW transformations at these defects are
caused by single electron impacts, we carried out atomistic
simulations of such impacts onto atoms near di-vacancies. Due to the
computational cost related to many non-equivalent atoms present in the
system and a large number of possible atomic configurations, we could
not repeat the detailed analysis of the role of initial space angle of
the displacement similar to pristine graphene, and therefore limited
our simulations to the $\theta = 0^\circ$ case for all non-equivalent
atoms at reconstructed di-vacancy structures. However, because the
defects break the symmetry of the lattice, a directional preference
for the displacements arises (atom displaced in the perpendicular
direction will change its direction due to local strain). This effect
is strong enough to facilitate bond rotations in reconstructed
di-vacancy structures. In Figure~\ref{px::mvacsim} we present examples of
such processes for both $V_2$(5-8-5)$\rightarrow V_2$(555-777) and
$V_2$(555-777)$\rightarrow V_2$(5555-6-7777) transformations.

Within these simulations, we never observed $V_2$(555-777)$\rightarrow
V_2$(5-8-5) transformations. Because the symmetry of the
$V_2$(555-777) defect around the middle atom is the same as that of
pristine graphene (the middle atom is represented as a black sphere in
the last panel of Figure~\ref{px::mvacsim}a), one would also expect that a
$\theta \neq 0^\circ$ displacement is required for this
transformation, similar to the {\sw} case, although in principle the
surrounding atoms could also cause this transformation. We noticed
that the most likely di-vacancy transformation, at least for
$\theta=0^\circ$, is the $V_2$(5555-6-7777)$\rightarrow
V_2$(555-777). The $V_2$(5-8-5)$\rightarrow V_2$(555-777) was the
least likely one of those observed. We never observed a
$V_2$(5-8-5)$\rightarrow 2\times$(5-7)
transformation~\cite{Kotakoski2011} during our simulations, which we
also attribute to the limited simulated conditions
($\theta=0^\circ$). Curiously, however, we did observe one
transformation in which a $V_2$(5-8-5) di-vacancy directly migrated
one step in the zigzag lattice direction.

Another interesting observation originating from these simulations is
the fact that the displacement threshold for atoms in the central part
of the reconstructed defects [$V_2$(555-777) and $V_2$(5555-6-7777)]
are higher than that for pristine graphene (by as much as 5\%). This
may explain why defect structures tend to grow into larger and larger
amorphous patches instead of collapsing into holes under continuous
electron irradiation at low voltages ($\leq
100$~keV)~\cite{Kotakoski2011}: Even when atoms are removed from the
defected area, the displacements occur at the edges of the existing
defects rather than at the central part where the local atomic density
is already lower. Clearly, since the core structure of these defects
consists of carbon hexagons, there must exist a limiting size above
which the displacement threshold becomes similar to that of ideal
graphene.

\section{Conclusions}

To conclude, by combining AC-HRTEM experiments and atomistic
simulations, we have shown that the bond rotations which lead to
creation of topological defects in carbon nanostructures are caused by
single electron impacts or incomplete annihilation of Frenkel
defects. This explains the discrepancy between experimental
observations of Stone-Wales defects and their relatively high
formation energy and even higher energy barrier for bond rotation.

The SW transformation in graphene can be initiated at least in two
different ways upon electron impact (involving a ``circling'' or
``nudging'' motion), and for almost any space angle, provided that
enough energy is transferred from the electron to the target atom.
Our simulations indicate that {\sw} can appear as a result of
``incomplete'' recombination of a Frenkel defect reminiscent of the
formation of Wigner-type defects in silicon~\cite{Tang1997}. However,
this is much more likely in the case of local curvature, as in
nanotubes. More surprisingly, we also observed sputtering of C$_2$
dimer as a result of annihilation of a Frenkel defect in carbon
structures with high curvature.

Moreover, we showed that the displacement threshold of atoms in the
central area of reconstructed defects is higher than that of pristine
graphene, by as much as 5\%, which explains why defected graphene
under low-energy electron irradiation ($\lesssim 100$~keV) tends to
turn graphene into a two-dimensional amorphous
structure~\cite{Kotakoski2011} instead of a perforated membrane. For
different di-vacancy structures, even displacements in direction
perpendicular to the graphene layer can initiate SW transformation and
thus local structural changes and defect migration.

Our results provide microscopic insight into the irradiation-induced
changes in the atomic structure of carbon nanosystems under electron
irradiation, and taking into account the interesting electronic
properties of defects associated with SW
transformations~\cite{Kotakoski2011,Lherbier2011,Appelhans2010},
may open new avenues for irradiation-mediated
engineering~\cite{Krasheninnikov2010}
of carbon nanostructures with next-generation electron microscopes.

\section{Acknowledgments}

We acknowledge financial support by the German Research Foundation
(DFG), the German Ministry of Science, Research and Arts (MWK) of the
state Baden-Wuerttemberg within the SALVE (sub angstrom low voltage
electron microscopy) project and Academy of Finland through several
projects. We also thank CSC, Espoo, Finland, for generous grants on
computer time.

\end{document}